\documentclass[aps,prl,superscriptaddress,showpacs,twocolumn]{revtex4-1}

\usepackage{graphicx,graphics,epsfig}   
\usepackage{dcolumn}    
\usepackage{bm}         
\usepackage{amsmath}    
\usepackage{verbatim}   
\usepackage{color}      
\usepackage{subfigure}  
\usepackage{times,natbib}
\usepackage{amsmath,amsfonts,amssymb,graphics,graphics,color,times}

\usepackage{latexsym}
\usepackage{amsmath}
\usepackage{amssymb}
\usepackage{amsfonts}
\usepackage{amsthm}
\usepackage{mathrsfs}
\usepackage{color,verbatim,graphics}
\usepackage{psfrag}
\DeclareMathAlphabet{\mathrsfs}{U}{rsfs}{m}{n}
\DeclareMathAlphabet{\mathpzc}{OT1}{pzc}{m}{it}
\DeclareMathAlphabet{\matheus}{U}{eus}{m}{n}
\DeclareMathAlphabet{\mathbbold}{U}{bbold}{m}{n}

\newcommand{\ba}{\begin{eqnarray}}
\newcommand{\ea}{\end{eqnarray}}
\newcommand{\ban}{\begin{eqnarray*}}
\newcommand{\ean}{\end{eqnarray*}}
\newcommand{\be}{\begin{equation}}
\newcommand{\ee}{\end{equation}}

\newcommand{\ket}[1]{|#1\rangle}
\newcommand{\bra}[1]{\langle#1|}

\newcommand{\one}{\mathbbold{1}}

\newcommand{\ie}{{\it{i.e.}}}
\newcommand{\etal}{{\it{et al.}}}

\begin{document}

\title{All quantum states useful for teleportation are nonlocal resources}

\author{Daniel Cavalcanti}
\affiliation{Centre for Quantum Technologies, University of
Singapore, 117542, Singapore}
\author{Antonio Ac\'in}
\affiliation{ICFO-Institut de Ciencies Fotoniques, 08860 Castelldefels (Barcelona), Spain}
\affiliation{ICREA--Instituci\'o Catalana de Recerca i Estudis Avan\c{c}ats, Lluis Companys 23, 08010 Barcelona,
Spain}
\author{Nicolas Brunner}
\affiliation{H.H. Wills Physics Laboratory, University of Bristol, Tyndall Avenue, Bristol, BS8 1TL, United Kingdom}
\affiliation{D\'epartement de Physique Th\'eorique, Universit\'e de Gen\`eve, 1211 Gen\`eve, Switzerland}
\author{Tam\'as V\'ertesi}
\affiliation{Institute of Nuclear Research of the Hungarian Academy of Sciences,
H-4001 Debrecen, P.O. Box 51, Hungary}

\begin{abstract}
Understanding the relation between the different forms of
inseparability in quantum mechanics is a longstanding problem in
the foundations of quantum theory and has implications for quantum
information processing. Here we make progress in this direction by
establishing a direct link between quantum teleportation and Bell
nonlocality. In particular, we show that all entangled states
which are useful for teleportation are nonlocal resources, i.e.
lead to deterministic violation of Bell's inequality. Our result
exploits the phenomenon of super-activation of quantum
nonlocality, recently proved by Palazuelos, and suggests that the latter might in fact be generic.
\end{abstract}

\maketitle

The fact that quantum mechanics is at odds with the principle of
locality, although once viewed as evidence of the failure of the
model, is nowadays recognized as a fundamental aspect of quantum
theory. The notion of inseparability in quantum mechanics appears
under several forms. At the algebraic level is
the concept of quantum entanglement. Entanglement is the resource
for numerous quantum information protocols, in particular for
quantum teleportation~\cite{bennett}, which plays a central role
in quantum communications and computation. The strongest notion of
quantum inseparability is Bell nonlocality \cite{bell}. Distant
observers sharing an entangled state can, by performing suitably
chosen local measurements, generate correlations that
cannot be reproduced by any local hidden-variable
model~\cite{EPR}, as witnessed by the violation of a Bell
inequality.

Understanding how these various forms of inseparability relate to
each other is long-standing problem, important from both a
fundamental and applied point of view. Although entanglement and
nonlocality were first thought to be equivalent, it was shown that
there exist mixed entangled states which are local, that is, admit
a local hidden variable model~\cite{werner}, even for the most
general type of local measurements~\cite{barrett}. It was then
shown by Popescu \cite{sandu2} that some of these local entangled
states are nevertheless useful for teleportation, which lead him
to argue that teleportation
and Bell nonlocality are inequivalent forms of inseparability.
Some time later, it was shown \cite{horo_teleport2} that all
two-qubit states violating the simplest Bell inequality,
due to Clauser-Horne-Shimony-Holt (CHSH), are useful for
teleportation, reviving the hope for a link between teleportation
and nonlocality.

Here we show that a strong link between teleportation and
nonlocality does indeed exist. In particular, we will show that
all entangled states useful for teleportation are nonlocal
resources, that is, can be used to violate a Bell inequality
deterministically. The key point in our work is to allow the
parties to use several copies of the state. More precisely, for
all states $\rho$ useful for teleportation (\ie states with
entanglement fraction higher than $1/d$ \cite{horo_teleport}),
there exists a finite number $k$ such that $k$ copies of
$\rho$, \ie $\rho^{\otimes k}$, deterministically violate a Bell.

The fact that, by combining several copies of a state that admits
a local model, it becomes possible to violate a Bell inequality is
known as super-activation of quantum nonlocality. This represents
a novel example of the phenomenon of activation in quantum
mechanics, through which the judicious combination of several
quantum entities becomes more powerful than the sum of the parts.
Celebrated examples of activation were demonstrated in
entanglement theory \cite{BEact} and quantum channel theory
\cite{smith}. Initial studies discovered examples of
super-activation of nonlocality in specific scenarios.
Refs.~\cite{sandu,peres1,gisin,masanes} discussed activation of
nonlocality in the case where the parties are allowed to locally
pre-process their state. Activation of nonlocality was also
discussed in the context of quantum networks
\cite{sen,dani,rabelo,grudka}, in the CHSH scenario
\cite{miguel,liang} and for general nonsignaling correlations
\cite{brunner}. More recently, Palazuelos \cite{Carlos} discovered
the first example of super-activation of quantum nonlocality in
the most natural scenario consisting of two parties who do
not perform any local pre-processing of their quantum state.
Central to his analysis, are Bell inequalities discussed by
Buhrman et al.~\cite{buhrman}, based on a game due to Khot and
Visnoi \cite{KV}, which represent one of the first explicit
examples of unbounded Bell inequality violations in quantum
mechanics.

Below, building upon the work of Ref.~\cite{Carlos}, we first show
that any isotropic state (i.e. a mixture of a maximally entangled
state and white noise) that is entangled can be super-activated.
In turn, we show that any bipartite entangled state in
$\mathbb{C}^d\otimes \mathbb{C}^d$ with entanglement fraction
larger than $1/d$ can be super-activated, thus establishing our
main result, i.e. direct link between quantum nonlocality and
quantum teleportation. Based on this result, we provide
new examples of activation of quantum non-locality. Finally, we
discuss perspectives for future research.

\emph{Superactivation of nonlocality.--} In Ref.~\cite{Carlos},
Palazuelos studied the nonlocal properties of many copies of an
isotropic state, given by \be\label{isotropic} \rho_{{\rm
iso}}(p)=p\ket{\Phi_d}\bra{\Phi_d}+(1-p)\frac{\one}{d^2}, \ee
where $0\leq p\leq 1$ is a noisy parameter,
$\ket{\Phi_d}=\sum_{i=1}^d\ket{ii}/\sqrt{d}$ is a maximally
entangled state of local dimension $d$ and $\one/d^2$ is the
$d\times d$ maximally mixed state. It is known that there exist
ranges of the noisy parameter $p_{{\rm sep}}\leq p\leq p_{{\rm
loc}}$ such that this state is local, \ie~ the probability
distributions of the results of local measurements applied to
\eqref{isotropic} can be explained by a (classical) local
model~\cite{Almeida}. However, Palazuelos showed that sufficiently
many copies of a local isotropic state, \ie~ $\rho_{{\rm
iso}}(p_{{\rm loc}})^{\otimes k}$, can violate a Bell inequality
if $d\geq8$ \cite{Carlos}. This shows that quantum nonlocality can
be super-activated.

Here we prove that every entangled isotropic state (in any
dimension) can be used to obtain nonlocality, in the sense that
many copies of it are nonlocal. To do that we will basically
follow the same proof as in~\cite{Carlos}.
The Bell inequality considered by
Palazuelos is based on the Khot-Visnoi Bell inequality \cite{KV},
\be\label{loc bound} \beta_{KV}=\sum_{x,y=1}^{N}\sum_{a,b=1}^{n}
c_{x,y,a,b}P(ab|xy)\leq \kappa_{{\rm loc}}, \ee where
$\kappa_{{\rm loc}}$ is the local bound and $c_{x,y,a,b}$ are
positive coefficients. The following upper bound is known for
$\kappa_{{\rm loc}}$: \be \kappa_{{\rm loc}}\leq C/n, \ee where
$N=2^n/n$ is the number of measurements per party, $n$ the number
of measurement outcomes \cite{footnote}, and $C$ a universal constant. Moreover,
it is known that there exist local measurements on a maximally
entangled state of local dimension $n$ which produce a probability
distribution such that \be\label{violation} \beta_{KV}\geq C'/(\ln
n)^2, \ee being $C'$ another universal constant. Since both the
local bounds and quantum violation depend on the dimension it is
useful to define the nonlocality fraction, given by \be
LF=\beta_{KV}/\kappa_{{\rm loc}}. \ee So, we have a nonlocal
probability distribution if $LF>1$. In what follows we sometimes
denote by $LF(\rho)$ the nonlocal fraction of the probability
distribution obtained by local measurements applied to the state
$\rho$.

\emph{Main Results.--} In what follows we say that a state $\rho$
is $k-$copy nonlocal if $\rho^{\otimes k}$ is nonlocal for some
$k$. We start by proving the following result.

\textbf{Result 1.--} Every entangled isotropic state is $k-$copy nonlocal.

For this sake, let us follow the proof of Palazuelos with the
isotropic state written in the (slightly different) following way:
\be\label{isotropic2} \rho_{{\rm
iso}}(F)=F\ket{\Phi_d}\bra{\Phi_d}+(1-F)\frac{\one-\ket{\Phi_d}\bra{\Phi_d}}{d^2-1},
\ee where $F=\bra{\Phi_d}\rho_{{\rm
iso}}(d)\ket{\Phi_d}=p+(1-p)/d^2$ is the entanglement fraction
\cite{Horodecki} of the state $\rho_{{\rm iso}}(F)$.

First consider $k$ copies of the isotropic state \be \rho_{{\rm
iso}}^{\otimes k}=F^k \ket{\Phi_{d^k}}\bra{\Phi_{d^k}}+...+(1-F)^k
\frac{\left(\one-\ket{\Phi_d}\bra{\Phi_d}\right)^{\otimes
k}}{(d^2-1)^k}, \ee where we have used that the tensor product of
$k$ maximally entangled states of local dimension $d$ is a
maximally entangled state of local dimension $d^k$. One can use
the bounds \eqref{loc bound} and \eqref{violation} to show that
the nonlocal fraction of $\rho_{{\rm iso}}^{\otimes k}$ satisfies
\ba\label{iso violation}
LF(\rho_{{\rm iso}}^{\otimes k})&=&F^k LF(\ket{\Phi_{d^k}}\bra{\Phi_{d^k}})+...  \nonumber \\
 & & +(1-F)^k LF\left(\frac{\left(\one-\ket{\Phi_d}\bra{\Phi_d}\right)^{\otimes k}}{(d^2-1)^k}\right)\nonumber\\
&\geq&F^k LF(\ket{\Phi_{d^k}}\bra{\Phi_{d^k}})\nonumber\\
&\geq&\frac{C'}{C}F^k\frac{d^k}{(k\ln d)^2}. \ea Thus, in the case
$F>1/d$ the right-hand-side of \eqref{iso violation} increases
with $k$. This implies that there will be a given number of copies
$k'$ for which $LF(\rho_{{\rm iso}}^{\otimes k'})>1$, which
implies that $LF(\rho_{{\rm iso}}^{\otimes k'})$ is nonlocal.
Since $F=1/d$ is the separability bound for the isotropic state we
conclude that every entangled isotropic state is $k-$copy
nonlocal.

\begin{figure}[tbp]
\includegraphics[width = 0.48\textwidth]{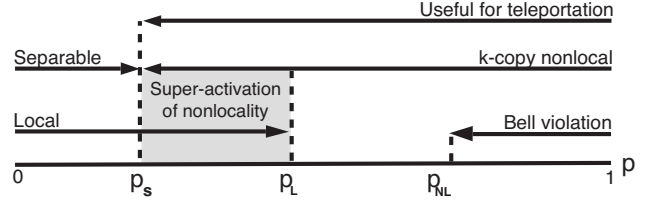}
\caption{Nonlocal properties of the isotropic state
\eqref{isotropic}. This state is separable for $p\leq
p_s=1/(d+1)$, has a local model for $p\leq p_L$ \cite{Almeida},
and is known to violate a Bell inequality for $p_L<p_{NL}<p$
\cite{CGLMP}. Here we show that several copies of the isotropic
state is nonlocal if it is entangled, that is, if $p_s<p$. The
state is useful for teleportation down to the separability limit
$p_s$.} \label{scheme}
\end{figure}

Result 1 extends the result by Palazuelos in two senses. First it
shows super-activation of nonlocality for the whole range of noisy
parameters for which the isotropic state is entangled. Second it
is valid for any dimension, while the minimum local dimension used
in the proof of Ref. \cite{Carlos} is $d\geq8$.


\textbf{Result 2.--}Every quantum state with entanglement fraction
higher than $1/d$ is $k-$copy nonlocal.


In order to see this, first notice that any quantum state $\rho_0$
can be depolarized into an isotropic state by the application of
randomly chosen unitaries as \be \rho_{{\rm iso}}(F_0)=\sum_i p_i
(U_i\otimes U_i^*) \rho_0 (U_i\otimes U_i^*)^{-1} \ee while
keeping the entanglement fraction $F_0$, \ie~
$F_0=\max_{\Psi}\bra{\Psi}\rho_0\ket{\Psi}$, where the maximum is
taken over all $d\times d$ maximally entangled states $\Psi$
\cite{Horodecki,Gross}. Since the initial state has
entanglement fraction higher than $1/d$, the resulting isotropic
state is entangled and, thus, $k$-copy non-local. That is, the
state \be\label{decomp} \rho_{{\rm iso}}(F_0)^{\otimes
k}=\sum_{i_1,...,i_k} p_{i_1} \cdots p_{i_k} \,
\mathcal{U}_{i_1,...,i_k} \, \rho^{\otimes k}_0 \,
\mathcal{U}_{i_1,...,i_k}^{-1} \ee where $\mathcal{U}_{i_1...i_k}=
(U_{i_1}\otimes ...\otimes U_{i_k})_{Alice} \otimes
(U^*_{i_1}\otimes ...\otimes U^*_{i_k})_{Bob}$, violates the
Khot-Visnoi Bell inequality. This means that there is a Bell
operator $\mathcal{B}$
such that $\text{tr}(\mathcal{B} \, \rho_{{\rm iso}}(F_0)^{\otimes
k})>L$ ($L$ being the local bound of the corresponding Bell
inequality). But then, there is (at least) one term in the sum of
\eqref{decomp} such that $\text{tr}(\mathcal{B} \,
\mathcal{U}_{i_1,...,i_k} \rho^{\otimes k}_0
\mathcal{U}_{i_1,...,i_k}^{-1} )>L$. Since the local unitaries can
be absorbed into the choice of local measurements,
it follows that the original state $\rho^{\otimes k}_0$ is also
$k$-copy nonlocal.

Result 2 can be used to construct new examples of
super-activation of nonlocality:

\textbf{Result 3.--} Every two-qubit entangled state of the form
\be\label{sigma} \sigma=p\ket{\psi}\bra{\psi}+(1-p)\frac{\one}{4}
\ee is $k-$copy nonlocal, where $\ket{\psi}$ is an arbitrary
two-qubit pure entangled state. There exist a range of the noise parameter $p$ for
which the state \eqref{sigma} is known to be entangled and
local~\cite{Almeida}. 

To see this, let us first consider an arbitrary bipartite pure
state $\ket{\psi}$ in its Schmidt decomposition:
$\ket{\psi}=\sum_{i=1}^d \lambda_i\ket{ii}$, so that its
entanglement fraction can be written as \be
F_\psi=|\bra{\Phi_d}\psi\rangle|^2=\frac{(\sum_{i=1}^d
\lambda_i)^2}{d}. \ee By Result 2, $\sigma$ is $k-$copy nonlocal
if
$F_\sigma=\bra{\Phi_d}\sigma\ket{\Phi_d}=pF_\phi+(1-p)/d^2>1/d$.
This implies that \be
 p>\frac{d-1}{d(\sum_{i=1}^d\lambda_i)^2-1}.
 \ee
 For $d=2$ this bound is exactly the separability bound for the state $\sigma$.


Another class of two-qubit states with a local model is the one
given by \be\label{almeida state} \rho = p\ket{\psi}\bra{\psi} +
(1-p) \rho_A \otimes \frac{\one}{2}, \ee where
$\ket{\psi}=\cos{\theta}\ket{00}+\sin{\theta}\ket{11}$
($0\leq\theta\leq\pi/4$) is an arbitrary two-qubit state with
reduced state $\rho_A$. This state has a local model for $p\leq
5/12$ \cite{Almeida}. Imposing that the entanglement fraction of
\eqref{almeida state} is bigger than 1/2 implies $p>
1/(1+2\sin(2\theta))$. Using the partial transpose criterion, the
separability limit is found to be $p_{sep}=1/3$. Hence the state
\eqref{almeida state} for all $0 < \theta \leq \pi/4$ when
$p>1/(1+2\sin(2\theta))$ is $k$-copy nonlocal, and there is
super-activation (considering $p_{{\rm loc}}=5/12$) above
$\theta>(1/2)\arcsin(7/10) \sim 0.3877$ rad.

Result 1 can also be used to show super-activation of nonlocality
in the multipartite scenario. In Ref. \cite{TothAcin} a genuinely
entangled multipartite state of three qubits with a local model
was presented. It turns out that the reduced states of this
tripartite state are entangled two-qubit isotropic states. Thus,
bipartite nonlocality can be obtained by $k$ copies of this
tripartite state.

\emph{Discussion.--} We have shown that all entangled states
useful for teleportation are nonlocal resources. Although some of
these states admit a local hidden variable model, a Bell
inequality is always violated deterministically when suitable
local measurements are performed on a sufficiently large number of
copies of the state, hence leading to the phenomenon of
super-activation of quantum nonlocality. Thus our result
establishes a direct and general link between teleportation and
nonlocality, two central forms of inseparability in quantum
mechanics previously thought to be unrelated \cite{sandu2}.

These results raise several important questions. First, it would
be interesting to see if the converse to the present result holds,
that is, if any state which violates a Bell inequality is useful
for teleportation. Indeed, such a link has been established for
the particular case of the CHSH Bell inequality
\cite{horo_teleport2}. If this is the case, then teleportation and
nonlocality would turn out to be equivalent. Moreover, Bell
inequality violation would represent a device-independent test for
usefulness in teleportation.

Second, our results imply that a large class of entangled states
can be super-activated. Thus it is natural to ask whether
super-activation is a fully generic phenomenon, that is: is any
entangled state $\rho$ a nonlocal resource when a sufficient
number of copies of $\rho$ is considered? Indeed, a positive
answer to this question would demonstrate a direct link between
entanglement and nonlocality, which would considerably improve our
understanding of these concepts.

We believe however that answering this question is highly
challenging. Nevertheless, partial answers would already represent
breakthroughs. For instance, is it the case that any entangled
state that is distillable can be super-activated? Note that our
result applies to states which are distillable, as an
entanglement fraction higher than $1/d$ is a sufficient condition
for entanglement distillability. This condition is however not
necessary and, therefore, there exist distillable states to which
our result does not apply to. More ambitious is the case of bound
entangled states, i.e. which cannot be distilled. If
super-activation was found to be possible for such states, this
would disprove a long-standing conjecture made by Peres
\cite{peres2}. More generally this would show that bipartite
nonlocality does not imply distillability of entanglement, a
result known to hold true in the multipartite setting \cite{VB}.

\emph{Acknowledgements.} We thank Y.-C. Liang and C. Palazuelos for useful comments.
This work was supported by the National Research Foundation and
the Ministry of Education of Singapore, the UK EPSRC, the Spanish
project FIS2010-14830 and Generalitat de Catalunya, by the
European PERCENT ERC Starting Grant, Q-Essence, QCS and DIQIP
Projects, the Hungarian National Research Fund OTKA
(PD101461), a J\'anos Bolyai Grant of the Hungarian Academy of
Sciences, and the Swiss National Science Foundation (grant PP00P2\_138917). DC and NB
thank AA for the hospitality at ICFO. DC thanks NB for the
hospitality at the University of Bristol.


\end{document}